\documentclass{ws-procs9x6}

\begin{document}

\title{Spin Physics Progress with the STAR Detector at RHIC}

\author{J. SOWINSKI}

\address{Indiana University Cyclotron Facility, \\
2401 Milo B. Sampson Ln., \\ 
Bloomington, IN 47408, USA\\ 
E-mail: sowinski@iucf.indiana.edu}

\author{for the STAR Collaboration}

\maketitle

\abstracts{
Progress in STAR over the last two years and projections for the coming two
years are discussed.  Important technical developments have been the 
completion of the barrel and endcap electromagnetic calorimeters.  Measurement
of inclusive jet and $\pi_0$ A$_{LL}$ over the next two years are expected
to answer whether $\Delta$G is a major contributor to the proton's spin.
Transverse effects such as Collins and Sivers functions at mid-rapidity will
also be investigated.
}

Since the first data from polarized proton collisions at RHIC were
presented\cite{rak} at SPIN 2002 the STAR collaboration has
continued to participate in the short data acquisition periods
provided as the beams were developed in 2003 and 2004.  During these
periods, characterized by increasing luminosity and beam polarization,
STAR has been able to commission new detectors important to its spin
program as well as take data that will help tune the detectors for the
upcoming runs and provide ``first looks'' with limited statistics at
spin observables necessary to understanding the spin structure of the
nucleon.  With luminosity and polarization continuing to
improve\cite{waldo}, it is expected that extended running periods
for polarized protons will be provided over the next few years
allowing significant constraints to be placed on $\Delta$G and
investigations of transverse physics such as Sivers and Collins
functions to be started.  In the following I will review STAR's
upgrades of the past few years, the measurements already made and
prospects for the next few years leading to SPIN 2006.

The STAR detector\cite{stnim} is based on a large solenoidal magnet
with much of its internal volume devoted to tracking charge particles
with a time projection chamber(TPC) in a 5 kG magnetic field.
A big effort has gone into providing
electromagnetic calorimetry around the outer diameter of the TPC, the
barrel electromagnetic calorimeter (BEMC)\cite{bemcnim}, and covering
one poletip of the magnet, the endcap electromagnetic calorimeter
(EEMC)\cite{eemcnim}.  Since 2002 subsets of these calorimeters
have been installed and commissioned, with the installation of the
final modules and electronics taking place in preparation for the 2005 run.
This completes STAR's large solid angle capabilities
allowing for the full reconstruction of jets (unique at RHIC) in
addition to providing fast triggers, both essential to the spin
program.

The two calorimeters are both lead/plastic-scintillator sampling
calorimeters capable of measuring gamma rays from below 1 GeV to over
100 GeV in energy.  The BEMC covers the range in pseudorapidity,
\break
$\eta$=~-ln(tan($\theta$/2)), of -1$\leq \eta \leq$1 ($\theta
\geq40^{\circ}$) with 4800 individually read out ``tower'' energy signals.
The EEMC provides an additional 720 tower energy signals in a forward
pseudorapidity range, 1.1$\leq \eta \leq$2 ($37^{\circ}\geq\theta
\geq 15^{\circ}$), important for reaching
small x$_{gluon}$ in partonic collisions.  The first two active layers
of each detector are read out twice, once as part of the tower sum and
independently as a pre-shower signal to help in $\pi^0$/$\gamma$ and
electron/hadron discrimination.  The last layer of the EEMC is also
read out twice, again in the tower sum and independently as a
post-shower signal for e/h discrimination.  Both detectors
incorporate a finely segmented shower maximum detector (SMD) approx. 5
radiation lengths deep in the detector.  The BEMC SMD is a gaseous
proportional wire chamber while the EEMC uses triangular scintillator
strips laminated into two crossed planes.  These SMDs are used to
reconstruct the 2 showers from $\pi ^0$ decay vs. the single shower of
direct photons.  Although not complete over the full solid angle, the
full functionality of these detector subsystems was used
in the 2004 RHIC run.  Currently algorithms such as 
those for finding $\pi ^0$s and jets are being fine tuned and physics analyses have begun.

It is well known that the spin of the quarks do not account for
the spin of a proton\cite{bunce}.  Other contributions such as the
spin of gluons ($\Delta$G) or the orbital angular momentum are poorly
constrained by existing data.  Partonic scattering processes in
$\vec{p}-\vec{p}$ scattering such as q + g $\rightarrow$ q + g, 
g~+~g$\rightarrow$~g~+~g and q + g $\rightarrow$ q + $\gamma$ detected as
jets, and in the last case an isolated direct photon, have large spin
sensitivities\cite{bunce}, opening a window to $\Delta$G.  The large
solid angle of STAR is particularly well suited for detecting jets and
direct-photon/jet coincidences over a range in pseudorapidity that
helps cover a broad range in the partonic momentum fraction x.

A long term goal for the STAR spin physics program is to measure
$\Delta$g(x), i.e. not just the integral $\Delta$G over some region
but the x dependence from 0.01$\le$x$\le$0.3, using the q + g
$\rightarrow$ q + $\gamma$ (QCD Compton scattering) process.  The
coincident direct-photon and jet present a relatively clean
experimental signal, the partonic subprocess dominates other partonic
subprocesses generating the same signal and two body kinematics allows
the reconstruction of the partons' x and scattering
angle.  However this is a rare process and will require over
100pb$^{-1}$ so that the  10s of pb$^{-1}$ expected in the next 2 runs
will serve primarily to tune the technique.

\begin{figure}[t]
\centerline{
\psfig{figure=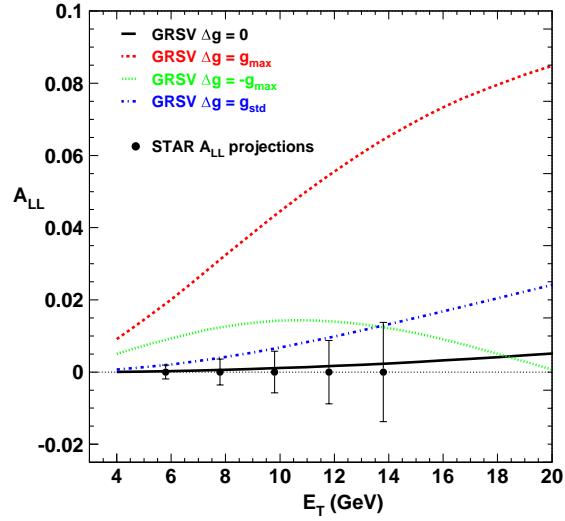,height=2.7in}
}
\caption{
Predictions of A$_L$$_L$ for inclusive jets$^7$
compared with the precision of measurements from data expected in
the 2005 running period.  An integrated luminosity of 7 pb$^-$$^1$ with
a polarization of 0.4 as well as jet finding efficiencies from
previous runs is assumed.
}

\label{fig:predict}
\end{figure}

Although not as selective, A$_{LL}$ measured with inclusive jets is
sensitive\cite{vogel} to $\Delta$G as shown in Fig. 1, and jets are produced
abundantly in the collisions at RHIC.  The STAR detector has a large
enough acceptance to reconstruct jets, and jet finding algorithms are
being tuned on the 2004 data set.  STAR triggers on jets either with a
high tower trigger (largest tower above a threshold) or jet patch
triggers (one of a predefined set of regions approx.  1$\times$1 in
$\eta\times\phi$ exceed a threshold).  These triggers were tested and used
during the 2004 run and preliminary analyses indicate that approximately 
1/2 of such triggers are found to contain jets with p$_T$ over 5~GeV.  It is
expected that when fully analyzed the 2004 data will only be able to
discriminate between the maximal $\Delta$G and 0.  The level of sensitivity
expected with 2005 data is displayed in Fig. 1.
The calorimeters in STAR are also capable
of reconstructing $\pi^0$s.  J\"ager et al. have shown\cite{vogel} that
jet sensitivity to $\Delta$G is carried over to the inclusive $\pi^0$
A$_{LL}$.  We expect that our inclusive $\pi^0$ results will have
similar discriminating power to that of the jets but with different
sensitivities to trigger bias.

A number of mechanisms have been suggested for explaining the forward
$\pi^0$ analyzing powers we have measured\cite{rak,akio}.  STAR is
also interested in pursuing effects arising from these at midrapidity.
If the asymmetries in jet decays reflecting the showering quark's
spin, as represented by Collins functions\cite{coll}, turn out to be
large enough, they can be used to gain access to transversity via spin
transfer from the final to initial state quarks.  Boer and Vogelsang
have also recently pointed out possible effects at a measurable level
resulting from gluonic Sivers functions in dijets.\cite{vogsiv} The
Sivers functions describe a correlation between the k$_T$(x) of
partons and the transverse spin of the proton.  While these
correlations integrate\cite{burk} to 0 over all partons and Bjorken x,
they need not be zero for individual partons.  STAR has already
measured the unpolarized k$_T$ distribution\cite{thomas} and it is of
sufficient width that, with transverse data expected in the next few
years, we should be able to observe signals of the predicted
size.\cite{vogsiv}

Improving polarization and luminosity as well as significant running
time for polarized protons at $\sqrt{s}$=200~GeV is expected over the
next few years.  In addition to continuing our studies of 
transverse physics, we hope that by spin 2006 we will be able to tell you
whether gluons are a major contributor to the proton's spin or not.

\end{document}